\documentclass[12pt,a4]{article}

\usepackage{amsmath}
\usepackage{amsfonts}
\usepackage{amssymb}
\usepackage{graphicx}

\def\be{\begin{eqnarray}}
\def\ee{\end{eqnarray}}

\def\h{{\rm  h}}

\def\q{{\bf q}}
\def\x{{\bf x}}
\def\k{{\bf k}}

\def\gi{{\rm gi}}

\def\cs2{c_{s}^{2}}

 \def\be   {\begin{equation}}   \def\ee   {\end{equation}}
 \def\ba   {\begin{array}}      \def\ea   {\end{array}}
 \def\bea  {\begin{eqnarray}}   \def\eea  {\end{eqnarray}}
 \def\bean {\begin{eqnarray*}}  \def\eean {\end{eqnarray*}}

\begin{document}

\begin{flushright} {\footnotesize CERN-PH-TH/2010-270\\DFPD-2010-A-18}  \end{flushright}
\vspace{5mm}
\vspace{0.5cm}
\begin{center}

\def\thefootnote{\fnsymbol{footnote}}

{\Large \bf The Gauge-Invariant Bias  of Dark Matter Haloes\\[0.2cm]
with  Primordial non-Gaussianity
}
\\[0.5cm]
{\large Nicola Bartolo$^{\rm a,b}$, Sabino Matarrese$^{\rm a,b}$ and 
 Antonio Riotto$^{\rm b,c}$}
\\[0.5cm]

{\small
\textit{$^{\rm a}$  Dipartimento di Fisica ``G. Galilei", Universit\`a degli Studi di Padova,\\ 
via Marzolo 8, I-35131 Padova, Italy}}

\vspace{.2cm}

{\small \textit{$^{\rm b}$  INFN Sezione di Padova, via Marzolo 8,
I-35131 Padova, Italy}}

\vspace{.2cm}

{\small \textit{$^{\rm c}$  CERN, PH-TH Division, CH-1211, 
Gen\`eve 23,  Switzerland}}

\end{center}

\vspace{.8cm}

\hrule \vspace{0.3cm}
{\small  \noindent \textbf{Abstract} \\[0.3cm]
The non-linear evolution of the halo population is  followed by 
solving the continuity equation 
under the  hypothesis that haloes move by the action of gravity. 
An exact and general formula for the  Eulerian bias field of dark matter haloes in terms of the Lagrangian bias is 
expanded at second-order  including the presence of primordial non-Gaussianity. Particular attention is paid
in defining  a gauge-invariant bias which is necessary when dealing with relativistic effects and
measured quantities. We show that 
scale-dependent effects in the Eulerian bias arise both at first- and second-order  independently from the presence
of some primordial non-Gaussianity. Furthermore, the Eulerian bias 
inherits from the primordial non-Gaussianity
not only a scale-dependence, 
but also a modulation with the angle of observation when sources with different biases are correlated. 
\vspace{0.5cm}  \hrule
\def\thefootnote{\arabic{footnote}}
\setcounter{footnote}{0}

\section{Introduction}
Cosmological inflation 
\cite{lrreview} has become the dominant paradigm to 
understand the initial conditions for the Cosmic Microwave Background (CMB) anisotropies
and Large Scale Structure (LSS) formation. 
This picture has recently received further spectacular confirmation 
by the Wilkinson Microwave 
Anisotropy Probe (WMAP) seven year set of data \cite{wmap5}.
Present and future \cite{planck} data 
may be sensitive to the non-linearities of the cosmological
perturbations at the level of second- or higher-order perturbation theory.
The detection of these non-linearities through the non-Gaussianity
(NG)  \cite{review} has become one of the primary experimental targets. 

A possible source of NG could be primordial in 
origin, being specific to a particular mechanism for the generation of the cosmological perturbations. This is what 
makes a positive detection of NG so
relevant: it might help in discriminating among competing scenarios which otherwise might be indistinguishable. Indeed,
various models of inflation, firmly rooted in modern 
particle physics theory, predict a significant amount of primordial
NG generated either during or immediately after inflation when the
comoving curvature perturbation becomes constant on super-horizon scales
\cite{review}. While standard single-field  \cite{noi}
and two(multi)-field \cite{two} models of inflation generically predict a tiny level of NG, 
`curvaton-type models' \cite{Enqvist:2001zp,Lyth:2001nq,Moroi:2001ct},  in which
a  significant contribution to the curvature perturbation is generated after
the end of slow-roll inflation by the perturbation in a field which has
a negligible effect on inflation, may predict a high level of NG \cite{luw}.
Alternatives to the curvaton model are those models 
characterized by the curvature perturbation being 
generated by an inhomogeneity in 
the decay rate \cite{hk,varcoupling} of the inflaton field. 
Other opportunities for generating the curvature perturbation occur
at the end of inflation \cite{endinflation} and during
preheating \cite{preheating}. All these models generate a level of NG which is local, since the NG part of the primordial curvature
perturbation is a local function of the Gaussian part generated on superhorizon scales. 
It has now become common to parametrize the level of NG through a dimensionless
quantity  $f_{\rm NL}$  which sets the 
magnitude of the three-point 
correlation function  \cite{review}.
In momentum space, the three point function (bispectrum), arising from the local NG is dominated by the
so-called ``squeezed'' configuration, where one of the momenta is much smaller than the other two and it is parametrized by
the non-linearity parameter $f_{\rm NL}^{\rm loc}$. Other models, such as DBI inflation
\cite{DBI} and ghost inflation \cite{ghost}, predict a different kind of primordial
NG, called ``equilateral'', because the three-point function for this kind of NG is peaked on equilateral configurations, 
in which the lengths 
of the three wave-vectors forming a triangle in Fourier
space are equal \cite{Shapes}. The equilateral NG is parametrized by an amplitude $f_{\rm NL}^{\rm equil}$~\cite{CN}. Present limits
on NG are summarized by $-10<f^{\rm loc}_{\rm NL}<74$ and   $-214<f^{\rm equil}_{\rm NL}<266$ at 95\% CL \cite{wmap5,Curto,zal}. 

It is clear that  detecting a significant amount of NG and its shape either from the CMB or from the
LSS offers the possibility of opening a window into the dynamics of the universe during the very first stages of its
evolution and to understand what mechanism gave rise to the cosmological perturbations.
Besides in the CMB anisotropies, NG  is particularly relevant in the high-mass end of
of density perturbations, i.e. on the scale of galaxy clusters, since the effect of NG fluctuations becomes especially visible on
the tail of the probability distribution function \cite{tail}. Furthermore, and more relevantly for us, 
primordial NG also alters the clustering of dark matter halos inducing a scale-dependent bias on large 
scales \cite{Dalal}. According to the peak-background split theory \cite{bardeen86} the underlying idea behind the generation of a local bias is that galaxies tend to form 
in regions where the dark matter density field is larger than some threshold value in Lagrangian space. The collapse of objects on small scales is ascribed to  the high 
frequency modes of the density fields, while the action of large-scale structures of these non-linear condensations is due to a shift of the local background density. 
As primordial NG generates a cross-talk between short and long wavelengths, it  alters significantly the local bias and introduces
a strong scale dependence in it. 
As a result, measuring the clustering properties of haloes is a  sensitive probe of primordial NG
which could be detected or significantly constrained by the various planned large-scale galaxy surveys,
both ground based (such as DES, PanSTARRS and LSST) and in space
(such as EUCLID and ADEPT) \cite{komatsuprop}. 

When analyzing the impact of NG onto the bias of dark matter haloes various points should be addressed. 
Being the effect computed at  second-order in perturbation theory one should consistently
calculate how to go from the Lagrangian bias to the Eulerian bias at the same order in perturbation theory. 
Furthermore, since the primordial NG manifests itself on large cosmological scales one should treat carefully the  
relativistic effects. This automatically calls for a gauge-invariant formulation of the
observables at hand. In this paper we address these points and show that a refined gauge-invariant treatment
of the Eulerian bias at second-order in perturbation theory leads to the prediction that
the bias is scale-dependent on large scales even in the absence of primordial NG and that the latter generates
an angular modulation  if sources with different biases are cross-correlated.

The paper is organized as follows. In section 2 we describe how to obtain the Eulerian
bias description in terms of the Lagrangian bias description at the non-perturbative level.
In section 3 we specificaly deal with the gauge-invariant formulation and in section 4 we proceed with the computation of the gauge-invariant Eulerian bias at second-order in perturbation theory. Finally,
section 5 contains our conclusions.

\section{Eulerian description of the bias from the local Lagrangian bias}
\noindent
In this section we describe how to obtain an expression for the Eulerian halo-to-mass bias 
starting from the local Lagrangian bias. According to the local Lagrangian description, the sites of the
galaxy formation are identified with specific regions of the primordial density field. The primordial galaxy
density field measuring the (smoothed) overdensity of galaxies in fieri at the Lagrangian position $\q$ at a given time
$\tau\ll 1$ is biased with respect to the primordial (linear) matter CDM density at the same location and at the same time. 
The Eulerian bias is obtained 
by integrating 
the continuity equations for the mass and  for the halo number density where galaxies are supposed to reside. 
The procedure follows and generalizes  the one provided in Ref. \cite{biasn} for the Newtonian case. 
We consider a spatially flat Universe filled with a cosmological constant 
$\Lambda$ and a non-relativistic pressureless fluid of Cold Dark Matter (CDM), 
whose energy-momentum tensor reads 
$T_{\mu\nu}=\rho u_{\mu} u_{\nu}$ where $u^\mu$ $(u_\mu u^\mu=-1)$ is the comoving four-velocity. 

Following the notations of Ref.~\cite{MMB}, the perturbed line element around 
a spatially flat FRW background reads
\begin{equation}
\label{metric}
ds^2=a^2(\tau)\{-(1+2\phi)d\tau^2 + 2 \hat{\omega}_i d\tau dx^i+[(1-2\psi) \delta_{ij} + \hat{\chi}_{ij}]dx^i dx^j \}\, .
\end{equation}
where $a(\tau)$ is the scale factor as a function of conformal time $\tau$. 
Here each perturbation quantity can be expanded into a 
first-order (linear) part and a second-order contribution, as for example, 
the gravitational potential $\phi=\phi^{(1)}+\phi^{(2)}/2$. 
Up to now we have not choosen any particular gauge. 
We can employ the standard split of the perturbations into the so-called scalar, 
vector and tensor parts,  
according to their transformation properties with respect to 
the $3$-dimensional space with metric $\delta_{ij}$, where scalar parts are 
related to a scalar potential, vector parts to transverse (divergence-free) 
vectors and tensor parts to transverse trace-free tensors. 
Thus $\phi$ and $\psi$, the gravitational potentials, are scalar perturbations, 
and for instance, $\hat{\omega}_i^{(r)}=\partial_i\omega^{(r)}+\omega_i^{(r)}$, 
where $\omega^{(r)}$ is the scalar part and $\omega^{(r)}_i$ 
is a transverse vector, {\it i.e.} $\partial^i\omega^{(r)}_i=0$ ($(r)=(1,2)$ stand for the 
$r$th-order of the perturbations). The symmetric traceless tensor $\hat{\chi}_{ij}$ generally contains a 
scalar, a vector and a tensor contribution, namely $\hat{\chi}_{ij} = D_{ij} \chi + 
\partial_i \chi_j + \partial_j \chi_i + \chi_{ij}$, where $D_{ij} \equiv \partial_i \partial_j - (1/3) \nabla^2 \delta_{ij}$,
$\chi_i$ is a solenoidal vector ($\partial^i \chi_i=0$) and  $\chi_{ij}$ represents a traceless and 
transverse (i.e. $\partial^i \chi_{ij} =0$) tensor mode\footnote{In what follows, for our purposes 
we will neglect linear vector modes since they are not produced in standard 
mechanisms for the generation of cosmological perturbations (as inflation), 
and we also neglect tensor modes at linear order, since they give a negligible 
contribution to LSS formation.}. 
The spatial projection tensor orthogonal to the fluid velocity
$u^\mu$ is defined by
\begin{equation}
h_{\mu\nu}=g_{\mu\nu}+u_\mu u_\nu, \quad \quad (h^{\mu}_{\nu} h^\nu_{\sigma}=h^\mu_{\sigma}, \quad h_\mu^{\nu}u_\nu=0).
\end{equation}
It is also useful to introduce the familiar decomposition
\begin{equation}
\nabla_\nu u_\mu=\sigma_{\mu\nu}+\omega_{\mu\nu}+{1\over 3}\Theta
h_{\mu\nu}-a_\mu u_\nu, \label{decomposition}
\end{equation}
where we have defined   the (symmetric) shear tensor $\sigma_{\mu\nu}$, the
(antisymmetric) vorticity  tensor $\omega_{\mu\nu}$, the volume expansion scalar $\Theta 
\equiv \nabla_\mu u^\mu$ and the acceleration $a_\mu\equiv u^\nu \nabla_\nu u_\mu$.
Notice that $\Theta$  reduces to
$3{\cal H}$ (${\cal H}$ being the Hubble rate in conformal time) in the homogeneous and isotropic
FRW case.

Our starting point is the 
conservation of the energy-momentum tensor of the CDM fluid,
\begin{equation}
\label{conserv}
\nabla_\mu T^\mu_{\nu}=0,
\end{equation}
which yields, after projecting along $u^\nu$, the continuity equation valid at any order in perturbation theory 
\begin{equation}
\label{continuity1} 
\dot\rho(\x,\tau) +  \Theta(\x,\tau)  \rho(\x,\tau) =0,
\end{equation}
where  the dot indicates  differentiation  along $u^\mu$, that is $\dot\rho=u^\mu\nabla_\mu\rho$.
If we now assume that a halo population of mass $M$ and formation time $\tau_{\rm f}$ is conserved
in time and evolves exclusively under the influence of gravity with an unbiased velocity\footnote{This approximation is accurate
if one is interested, as we are, on the bias at large-scales \cite{v}.}, 
meaning that the CDM fluid and haloes are moving with the same velocity, its number density $\rho_{\rm h}(\x,\tau)=
\rho_{\rm h}(\x,\tau | M,\tau_{\rm f})$ has to satisfy  the continuity equation

\begin{equation}
\label{continuity2}
\dot\rho_{\rm h}(\x,\tau)  + \Theta \rho_{\rm h}(\x,\tau)=0.
\end{equation}
Notice that Eqs. (\ref{continuity1}) and (\ref{continuity2}) are non-perturbative and valid in any
gauge. This is welcome as we want to derive the Eulerian bias factor which,  
being a physical observable,  must be a gauge-independent quantity. While comparing
the theoretical predictions (the matter power spectrum) with obervations (the galaxy
power spectrum)  does not represent a problem on sub-horizon scales where the matter density perturbations
computed in the different gauges all coincide, it is a delicate operation on scales comparable
with the horizon where different gauges provide different results even at the linear level~(see, e.g.,~\cite{YF}).
Truly gauge-independent perturbations must be
exactly constant in the background spacetime. This apparently
limits ones ability to make a gauge-invariant study of
quantities that evolve in the background spacetime, {\it e.g.}
density perturbations in an expanding cosmology.
In practice one can construct gauge-invariant definitions of unambiguous, that is physically defined, perturbations (see, e.g., the discussion of Ref.~\cite{MW}). These are not unique gauge-independent perturbations, but are gauge-invariant in the sense commonly
used by cosmologists to define a physical perturbation.
There is a  distinction  between quantities that are
automatically gauge-independent, {\it i.e.}, those that have
no gauge dependence (such as perturbations about a
constant scalar field), and quantities that are in general gauge-dependent (such as the curvature perturbation) but can have a gauge-invariant definition once their
gauge-dependence is fixed (such as the curvature 
perturbation on uniform-density hypersurfaces). In other words, one can define 
gauge-invariant quantities which 
are simply a coordinate independent definition of the
perturbations in the given   gauge. This can be often achieved by defining unambiguously a specific slicing into spatial hypersurfaces.
In this sense it should be clear that one may define an infinite number of, {\it e.g.},  gauge-invariant density contrasts. Which one to use
is a matter that can be decided only considering how the  determination of a given
observable   is performed. We will come back to this point later. For the time being it suffices to say that,
when expanded at a given order in perturbation theory, Eqs. (\ref{continuity1}) and (\ref{continuity2}) 
may be used to find the evolution of the  gauge-independent (in the sense just described) CDM and halo density contrasts, $\delta^\gi$ and $\delta_\h^\gi$.
These quantities  evolve following the same dynamics. This means that
their density contrasts $\delta^\gi$ and $\delta^\gi_\h$ at a given time will be related to their values at some initial time $\tau_{\rm in}\ll 1$ through the relation

\begin{equation}
\label {fund1}
\frac{1+\delta^\gi_\h(\x,\tau)}{1+\delta^\gi_\h(\q)}=
\frac{1+\delta^\gi(\x,\tau)}{1+\delta^\gi(\q)}.
\end{equation}
In Eq. (\ref{fund1}) by 
$\delta^\gi_\h(\q)=\delta^\gi_\h(\q| M,\tau_{\rm f})$ we mean the initial (Lagrangian) halo density fluctuation
at some primordial time  when 
 $\x(\tau=\tau_{\rm in})=\q$. 
Expanding Eq. (\ref{fund1}) up to second-order and setting   $\delta_\h=\delta_\h^{(1)}+\frac{1}{2}\delta_\h^{(2)}$,  we find
\begin{equation}
\label{a1}
\delta^{\gi(1)}_\h(\x,\tau)\simeq\delta_\h^{\gi(1)}(\q)+\left(\delta^{\gi(1)}(\x,\tau)-\delta^{\gi(1)}(\q)\right),
\end{equation}
and

\begin{eqnarray}
\label{a2}
\frac{1}{2}\delta^{\gi(2)}_\h(\x,\tau)&\simeq&
\frac{1}{2}\delta_\h^{\gi(2)}(\q)+
\frac{1}{2}\left(\delta^{\gi(2)}(\x,\tau)-\delta^{\gi(2)}(\q)\right)\nonumber\\
&+&\left(\delta_\h^{\gi(1)}(\q)-\delta^{\gi(1)}(\q)\right)\left(\delta^{\gi(1)}(\x,\tau)-\delta^{\gi(1)}(\q)\right),\nonumber\\
\end{eqnarray}
The expressions (\ref{a1}) and (\ref{a2})   are the key relations which permit to relate the Lagrangian bias to the Eulerian one. 
Before continuing though, we come back to the issue of which gauge-invariant contrasts we should take. 

\section{On the gauge-invariant formulation}
As we pointed out  before, there is an infinite number of ways to define gauge-invariant density contrasts  which differ by
other gauge-invariant combinations. Since one observes galaxies rather than the underlying matter
distribution and the latter at the source galaxy position is related to the mean matter density
at the observed redshift $z$, a good choice to define gauge-invariant density constrasts related
to each other by a bias factor 
seems the one involving the observed redshift $z$ \cite{YF}. At first order a coordinate transformation reads
$x^\mu\rightarrow x^\mu-\xi^\mu_{(1)}$ where $\xi^\mu_{(1)}=(\alpha^{(1)},
\xi^{(1)i})$. The matter density contrast 
transforms as $\delta^{(1)}\rightarrow \delta^{(1)}+\dot{\bar\rho}/\bar\rho \,\alpha^{(1)}$, where now dot stands for differentation with respect to the conformal time and  $\bar\rho\sim (1+z)^3$ is the background matter energy density; 
similarly the first-order perturbation of the
observed redshift transforms as $z^{(1)} \rightarrow z^{(1)} +\dot z\,\alpha^{(1)}$ (here $z$ is the unperturbed redshift). 
Going to the uniform redshift gauge  where the linear perturbation of redshift vanishes  relates  $\alpha^{(1)}$ to the linear perturbation of 
redshift in the old gauge, $\alpha^{(1)}=-z^{(1)}/\dot z$.
Therefore
the gauge invariant definition of the matter density contrast (and similarly for the halo one) is \cite{YF}

\begin{equation}
\label{d1}
\delta^{\gi(1)}=\delta^{(1)}-3\frac{z^{(1)}}{1+z}.
\end{equation}
At second-order the procedure is more involved, but straightforward. The coordinate transformation
reads $x^\mu\rightarrow x^\mu+\xi^\mu_{(1)}+\frac{1}{2}\left(\xi^\mu_{(1),\nu}\xi^\nu_{(1)}+
\xi^\mu_{(2)}\right)$ where $\xi^\mu_{(2)}=(\alpha^{(2)},
\xi^{(2)i})$. Under this
coordinate transformation the density matter  contrast and the redshift perturbation   transform as

\begin{eqnarray}
\delta^{(2)}&\rightarrow& \delta^{(2)}+\frac{\dot{\bar \rho}}{\bar\rho}\alpha^{(2)}+\alpha^{(1)}\left(\frac{\ddot{\bar \rho}}{\bar \rho}
\alpha^{(1)}+\frac{\dot{\bar \rho}}{\bar\rho}\dot{\alpha}^{(1)}+2\frac{\delta^{(1)}\dot \rho}{\bar\rho}\right)+\xi_{(1)}^i\left(\frac{\dot{\bar \rho}}{\bar\rho}\partial_i\alpha^{(1)}+2\frac{\partial_i \delta^{(1)}\rho}{\bar\rho}\right),\nonumber\\
z^{(2)}&\rightarrow& z^{(2)}+\dot{z}\alpha^{(2)}+\alpha^{(1)}\left(\ddot z
\alpha^{(1)}+\dot z\dot{\alpha}^{(1)}+2\dot{z}^{(1)}\right)+\xi_{(1)}^i\left(\dot{z}
\partial_i\alpha^{(1)}+2\partial_i z^{(1)}\right).
\end{eqnarray}
Going to the uniform redshift gauge where second-order perturbation of the redshift vanishes gives

\begin{equation}
\alpha^{(2)}=-\frac{z^{(2)}}{\dot z}+\frac{z^{(1)}\dot{z}^{(1)}}{\dot{z}^2}-\frac{\xi^i_{(1)}}{\dot z}\partial_i z^{(1)}.
\end{equation}
To completely solve the uniform redshift gauge at   second-order  we 
must also specify the first-order spatial gauge shift $\xi^i_{(1)}$. A natural choice is to pick 
worldlines comoving with the fluid. The (scalar) velocity transforms as $v_i^{(1)}\rightarrow
v_i^{(1)}-\xi_i^{(1)'}$. Thus from an arbitrary spatial gauge we can transform to
the comoving gauge by the spatial gauge transformation $\xi_i^{(1)}=\int v_i^{(1)}d\tau$.
Such a choice leads to the second-order gauge invariant matter density constrast (and similarly
for the halo one)

\begin{eqnarray}
\label{zz}
\frac{1}{2}\delta^{\gi(2)}&=&\frac{1}{2}\left(\delta^{(2)}-3\frac{z^{(2)}}{1+z}\right)+3\frac{z^{(1)}\dot{z}^{(1)}}{\dot z(1+z)}
+3\frac{\left(z^{(1)}\right)^2}{(1+z)^2}-\frac{z^{(1)}}{\dot z}\dot{\delta}^{(1)}\nonumber\\
&-&3\frac{z^{(1)}\delta^{(1)}}{1+z}
+\left(\partial_i\delta^{(1)}-3\frac{\partial_i z^{(1)}}{1+z}\right)\int d\tau v_i^{(1)}.
\end{eqnarray}
The next step amounts to determing the expression of the redshift perturbation in terms of the
perturbations of the metric (\ref{metric}) and other quantities.
Photons suffer a 
redshift $z$ during their travel from the emitter $\cal{E}$ to the 
observer $\cal{O}$;  the emitted frequency $\omega_{\cal{E}}$ and the 
observed one $\omega_{\cal{O}}$ are related by $\omega_{\cal{O}}=
\omega_{\cal{E}}/(1+z)$. Here $\omega=-g_{\mu\nu} u^\mu k^\nu$, where $u^\mu$ is the 
four-velocity of the observer or emitter and $k^\nu=dx^\nu/d\lambda$
is the wave vector of the photon in the conformal metric, tangent to 
the null geodesic $x^\nu(\lambda)$ ($\lambda$ is the affine parameter) followed by the photon from the
emission to the observation point. We do not report the full calculation  of the 
redshift perturbation which  basically amounts to solving for the photon trajectory. The computation can be found in Ref. \cite{MM}. Expanding the frequency as $\omega=\bar\omega(1+\omega^{(1)}+\frac{1}{2}\omega^{(2)})$, at first-order one obtains

\begin{eqnarray}
\label{zpert1}
\frac{z^{(1)}}{1+z}&=&\omega_{\cal E}^{(1)}-\omega_{\cal O}^{(1)}=
\phi^{(1)}_{\cal{O}}-\phi^{(1)}_{\cal{E}}+
v^{(1)i}_{\cal{E}} e_i-v^{(1)i}_{{\cal{O}}} e_i
+I_1(\lambda_{\cal{E}}),\nonumber\\
I_1(\lambda_{\cal{E}})&=&\int_{\lambda_{\cal{O}}}^{\lambda_{\cal{E}}}
d\lambda \,\dot{A}^{(1)},\nonumber\\
A^{(1)}&=& \psi^{(1)}+\phi^{(1)}+\hat{\omega}^{(1)}_i e^i-
\frac{1}{2}\chi^{(1)}_{ij}e^i e^j,
\end{eqnarray}
where $e^i$ indicates the zero-th order three-dimensional vector indicating the photon direction from which they arrive at the observer
~$\mathcal O$.
In the expression above one recognizes the Sachs-Wolfe effect due to the change in the gravitational potential at the source' and observer's points, the Doppler contribution due to the peculiar velocities
of the emitter and the observer and the integrated Sachs-Wolfe effect along the 
photon trajectory.

 At second-order expressions are more involved and the  perturbation of the redshift may be
written as \cite{MM}

\begin{eqnarray}
\label{z2}
\frac{1}{2}\frac{z^{(2)}}{1+z}&=&\frac{1}{2}
\left[\omega_{\cal E}^{(2)}-\omega_{\cal O}^{(2)}
-2 \omega_{\cal E}^{(1)}\omega_{\cal O}^{(1)}+2 \left(\omega_{\cal O}^{(1)}\right)^2\right]\nonumber\\
&=&
\frac{1}{2}\phi^{(2)}_{\cal{O}}
-\frac{1}{2}\phi^{(2)}_{\cal{E}}+
\frac{1}{2}v^{(2)i}_{\cal{E}} e_i-\frac{1}{2}v^{(2)i}_{{\cal{O}}} e_i
+\frac{3}{2}(\phi^{(1)}_{\cal{E}})^2
-\frac{1}{2}(\phi_{\cal{O}}^{(1)})^2-\phi^{(1)}_{\cal{O}}\phi^{(1)}_{\cal{E}}
+I_2(\lambda_{\cal{E}})
+v^{(1)i}_{\cal{E}} e_i \phi^{(1)}_{\cal{E}}\nonumber\\
&-&\left(I_1(\lambda_{\cal{E}})+v^{(1)i}_{\cal{E}} e_i\right)
\left(2 \phi^{(1)}_{\cal{O}}
-\psi^{(1)}_{\cal{O}}+\frac{1}{2}\chi^{(1)ij}_{\cal{O}}e_i e_j
-v^{(1)i}_{\cal{O}} e_i
-\phi^{(1)}_{\cal{E}}+v^{(1)i}_{\cal{E}} e_i
+I_1(\lambda_{\cal{E}})\right)\nonumber\\
&-&x^{(1)0}_{\cal{E}} \dot{A}^{(1)}_{\cal{E}}-(x^{(1)j}_{\cal{E}}
+x^{(1)0}_{\cal{E}} e^j)\left(\phi^{(1)}_{,j}-v^{(1)}_{i,j} e^i\right)_{\cal{E}}-v^{(1)i}_{\cal{O}}
\left(\frac{1}{2}v^{(1)}_{{\cal{O}}i}-2 \psi^{(1)}_{\cal{O}} e_i
+\chi^{(1)}_{{\cal{O}}ij} e^j\right)\nonumber\\
&+&\frac{1}{2}v^{(1)}_{{\cal{E}}i} v^{(1)i}_{\cal{E}}
+v^{(1)i}_{\cal{O}} e_i\left(\phi^{(1)}_{\cal{O}}
-\psi^{(1)}_{\cal{O}}+\frac{1}{2}\chi^{(1)kj}_{\cal{O}}e_k e_j
-\phi^{(1)}_{\cal{E}}\right)
\nonumber\\
&-&v^{(1)}_{{\cal{E}}i}\left(-\hat{\omega}^{(1)i}_{\cal{E}}
+\hat{\omega}^{(1)i}_{\cal{O}}+2 \psi^{(1)}_{\cal{O}}e^i
-\chi^{(1)ij}_{\cal{O}} e_j- I_1^i(\lambda_{\cal{E}})\right)\nonumber\\
&+&\left(v^{(1)i}_{\cal{E}} e_i-v^{(1)i}_{{\cal{O}}} e_i
+I_1(\lambda_{\cal{E}})\right)^2+2\left(\phi^{(1)}_{\cal{O}}-\phi^{(1)}_{\cal{E}}\right)\left(v^{(1)i}_{\cal{E}} e_i-v^{(1)i}_{{\cal{O}}} e_i
+I_1(\lambda_{\cal{E}})\right),
\end{eqnarray}
where
\begin{eqnarray}
I^i_1(\lambda_{\cal{E}})&=&\int_{\lambda_{\cal{O}}}^{\lambda_{\cal{E}}}
d\lambda \,{A}^{(1),i},\nonumber\\
I_2(\lambda_{\cal{E}})&=&\int_{\lambda_{\cal{O}}}^{\lambda_{\cal{E}}}
d\lambda \left[\frac{1}{2}\dot{A}^{(2)}-(\dot{\hat{\omega}}^{(1)}_i-\dot{\chi}^{(1)}_{ij} e^j)
(k^{(1)i}+e^i k^{(1)0})
+ 2 k^{(1)0} \dot{A}^{(1)}\right.\nonumber\\
&+&\left. 2 \dot{\psi}^{(1)} {A}^{(1)}
+x^{(1)0} \ddot{A}^{(1)}+x^{(1)i} \dot{A}^{(1)}_{,i}\right],\nonumber\\
A^{(2)}&=& \psi^{(2)}+\phi^{(2)}+\hat{\omega}^{(2)}_i e^i-
\frac{1}{2}\chi^{(2)}_{ij}e^i e^j
\end{eqnarray}
and
\begin{eqnarray}
k^{(1)0}(\lambda_{\cal{E}})&=&\phi^{(1)}_{\cal{O}}
-\psi^{(1)}_{\cal{O}}+\frac{1}{2}
\chi^{(1)ij}_{\cal{O}}e_i e_j-2 \phi^{(1)}_{\cal{E}}
-\hat{\omega}^{(1)i}_{\cal{E}} e_i +I_1(\lambda_{\cal{E}}),\nonumber\\
k^{(1)i}(\lambda_{\cal{E}})&=&2 \psi^{(1)}_{\cal{O}} e^i+
\hat{\omega}^{(1)i}_{\cal{O}}-\chi^{(1)ij}_{\cal{O}} e_j
-2 \psi^{(1)}_{\cal{E}} e^i-
\hat{\omega}^{(1)i}_{\cal{E}}+\chi^{(1)ij}_{\cal{E}} e_j
- I_1^i(\lambda_{\cal{E}}),\nonumber\\
x^{(1)0}(\lambda_{\cal{E}})&=&(\lambda_{\cal{E}}-\lambda_{\cal{O}})
\left[\phi^{(1)}_{\cal{O}}-\psi^{(1)}_{\cal{O}}+\frac{1}{2}
\chi^{(1)ij}_{\cal{O}}e_i e_j\right]+
\int_{\lambda_{\cal{O}}}^{\lambda_{\cal{E}}}d\lambda
\left[-2 \phi^{(1)}-\hat{\omega}^{(1)}_i e^i+(\lambda_{\cal{E}}-\lambda)
\dot{A}^{(1)}\right],\nonumber\\
x^{(1)i}(\lambda_{\cal{E}})&=&(\lambda_{\cal{E}}-\lambda_{\cal{O}})
\left[2 \psi^{(1)}_{\cal{O}} e^i+
\hat{\omega}^{(1)i}_{\cal{O}}-\chi^{(1)ij}_{\cal{O}} e_j\right]\nonumber\\
&-&\int_{\lambda_{\cal{O}}}^{\lambda_{\cal{E}}}d\lambda
\left[2 \psi^{(1)} e^i+\hat{\omega}^{(1)i}-\chi^{(1)ij} e_j
+(\lambda_{\cal{E}}-\lambda)A^{(1),i}\right] .
\end{eqnarray}
These long expressions  allow the determination of the gauge-independent density contrasts. 
In practice, one evaluates them adopting  the most appropriate gauge.
 One possible convenient choice for the determination of the Eulerian bias from the local Lagrangian bias 
is represented by the comoving-orthogonal gauge as we 
will explain in the following. Let us stress that the gauge-invariant expressions found in this subsection would be also useful
if, instead of computing the Eulerian bias derived from the local Lagrangian bias, one adopts the 
bias model described in Ref. \cite{fg}. In this approach the (smoothed) galaxy number density field at a given position $\x$ and time $\tau$ is assumed
to be a local function of the (smoothed) underlying CDM mass density at the same location and instant, $\delta_\h^\gi=b_1^{\rm E}\delta^\gi+\frac{1}{2}b_2^{\rm E}
\left(\delta^\gi\right)^2+\cdots$. This approach is essentially phenomenological and it is a priori devoid of any insight about the 
dynamics of the clustering. 

\section{The computation of the gauge-invariant Eulerian bias}
\noindent
In this section we proceed with the computation of the Eulerian bias adopting the gauge-invariant
formulation of the matter density perturbation described previously. It is convenient to perform such 
a computation collapsing to the comoving-orthogonal gauge. 
Indeed, as we  restrict ourselves to the case  of irrotational dust plus $\Lambda$,  
the comoving-orthogonal gauge  is also synchronous. 
Indeed, the fluid four-velocity field can be written as     $u^{\mu}=(1/a,0,0,0)$, so that ${\bf q}$ represents 
the comoving Lagrangian coordinate for the fluid element. The possibility of making the synchronous, time-orthogonal 
gauge choice and comoving gauge choice simultaneously is a peculiarity of fluids with vanishing spatial pressure gradients, i.e. 
vanishing acceleration, which holds at any time, {\it i.e.} also beyond the linear regime. 

The choice of the comoving-orthogonal  gauge to evaluate the  gauge-invariant 
 Eulerian bias   is motivated by various reasons. 
A simple analytic model for the gravitational clustering of dark matter haloes to understand how their spatial distribution is biased relative to that of the mass was developed in Ref. \cite{biasp}. The statistical distribution of dark haloes within the initial density field (assumed Gaussian) is determined by an extension of the Press-Schechter formalism and is done 
therefore at the Lagrangian level.  One then expects that the  gauge-invariant Eulerian description
to be therefore simpler to formulate through the synchronous gauge. Furthermore, 
the non-linear spherical collapse description necessary to compute the 
halo mass function through the (extended) Press-Schechter approach,   requires the choice of the 
comoving-orthogonal gauge.

In the synchronous and comoving gauge the line element reads
\begin{equation}
\label{oo}
ds^2=a^2(\tau)\left[-d^2\tau + h_{ij}({\bf x},\tau) dx^i dx^j\right]\, .
\end{equation}
Working in the synchronous comoving gauge,  the spatial coordinate 
does not evolve with time, $\x(\tau)=\x(\tau=\tau_{\rm in})\equiv \q$. 
From Eqs. (\ref{a1}) and (\ref{a2})
evaluated now 
in the comoving-orthogonal gauge, we may deduce the gauge-invariant Eulerian bias parameters. 
To
simplify our expressions we  limit ourselves to the case of a pure CDM-dominated universe.
One can easily check that the gauge-invariant matter density contrast (\ref{d1}) expressed
in the comoving-orthogonal gauge is given  by\footnote{To solve for the integral appearing in Eq.~(\ref{zpert1})
one can use, e.g., the results contained in the Appendix of Ref.~\cite{BarausseMR}. One can also readily check that the same
expression is obtained, {\it e.g.}, in the Poisson gauge.}
\begin{equation}
\delta^{\gi(1)}(\q,\tau)=
\frac{2}{3{\cal H}^2\Omega_m}\nabla^2\varphi(\q)+\varphi(\q)+\tau e^i\varphi_{,i}(\bf q),
\end{equation}
where $\varphi(\q)$ is the peculiar scalar gravitational  potential and we have got rid of the terms defined at the observer's point which
may be absorbed in the monopole term. Notice that in Fourier space the time evolution of the 
gauge-invariant density contrast as 

\begin{eqnarray}
\label{ev}
\delta^{\gi(1)}({\bf k},\tau)&=&f_\gi(\k,\tau)\delta^{(1)}_c(\k,\tau),\nonumber\\
f_\gi(\k,\tau)&=&1-\frac{6}{k^2\tau^2}-6i\frac{\hat{\k}\cdot{\bf n}}{k\tau},
\end{eqnarray}
where $\delta^{(1)}_c(\k,\tau)=-\frac{k^2\tau^2}{6}\varphi(\k)$ is the matter density contrast in the synchronous gauge.
The key point  is now to calculate the halo field $\delta^{\gi}_\h(\q)$. 
 The approach to the clustering evolution is based on a generalization of the so-called peak-background split, first proposed in Ref. \cite{bardeen86} which basically consists in splitting the mass perturbations in a fine-grained (peak) component $\delta^{\gi}_{\rm pk}$ 
filtered on a scale $R$ and a
coarse-grained (background) component $\delta^{\gi}_{\rm bg}$ filtered on a scale $R_0\gg R$. The underlying idea is to ascribe the collapse of objects on small scales to the high frequency modes of the density fields, while the action of large-scale structures of these non-linear condensations is due to a shift of the local background density. By definition, the Lagrangian distribution of nascent haloes of mass $M$ and formation time $\tau_{\rm f}$ is given  by 

\begin{equation}
\delta_\h^{\gi}(\q|M,\tau_{\rm f})\equiv\lim_{\tau\rightarrow \tau_{\rm in}}
b^{\rm E}(\q,\tau|M,\tau_{\rm f})\delta^{\gi}_{\rm bg}(\q,\tau)\equiv b^{\rm L}_{0}(\q|M,\tau_{\rm f})
\delta^{\gi}_{0}(\q),
\end{equation}
where $b^{\rm L}_{0}(\q|M,\tau_{\rm f})$ is the Lagrangian halo bias and  
$\delta^{\gi}_{0}(\q)$ is the mass density fluctuation linearly extrapolated to the present time $\tau_0$ (where $a(\tau_0)=1$) 
and filtered
on the background scale $R_0$; in Fourier space, $\delta^{\gi}_{0}(\k)=\left(
f_\gi(\k,\tau_0)/f_\gi(\k,\tau_{\rm in})a(\tau_{\rm in})\right)\delta^{\gi(1)}(\k,\tau_{\rm in})$.
One has to recall that the second equality in the latter equation does not mean that 
$\delta_\h^{\gi}(\q)$ is proportional to $\delta^{\gi}_{0}(\q)$. Indeed, the Lagrangian bias is in general a functional of the background density field. To understand the above equation, one has to recall that, 
 at sufficiently early times, the expression for the Eulerian bias field obtained in linear perturbation theory becomes exact (as the linear theory gets more and more accurate) and, going to Fourier space, 
 
 \begin{equation}
 \delta^{\gi}(\k,\tau)\simeq  \delta^{\gi}_{\rm bg}(\k,\tau)=\frac{f_\gi(\k,\tau)}{f_\gi(\k,\tau_0)}a(\tau)\delta^{\gi}_{0}(\k)
 \end{equation}
 when $\tau\simeq\tau_{\rm in}\ll 1$.  Notice that the perturbative expansion of  the matter density contrast 
is valid at sufficiently early times and/or large scales,  while 
the validity of the expansion of  the halo density contrast in powers of $\delta^{\gi}_{0}$ is based on assuming a suitably large smoothing radius $R_0$. Assuming these assumptions to hold and 
 expanding $\delta_{\rm h}^{\gi}(\q)$
  in powers of $\delta^{\gi}_{0}(\q)$

  \begin{equation}
  \delta_{\rm h}^{\gi}(\q) = \sum_{\ell\geq 1}\frac{b_{0\ell}^{\rm L}(M,\tau_{\rm f})}{\ell !}\left(\delta^{\gi}_{0}(\q)\right)^\ell,
  \end{equation}
    from Eq. (\ref{a1}) we  deduce that at first-order and  in at space 
  
 \begin{eqnarray}
 \label{bE1st}
\delta_\h^{\gi(1)}(\k,\tau)&\equiv&b_1^{\rm E}(\k,\tau)\delta^{\gi(1)}(\k,\tau)\simeq
(1+b_1^{\rm L}(\k,\tau))\delta^{\gi(1)}(\k,\tau),\\
b_1^{\rm L}(\k,\tau)&\simeq &\frac{f_\gi(\k,\tau_0)}{f_\gi(\k,\tau)}\frac{b^{\rm L}_{01}(M,\tau_{\rm f})}{a(\tau)}\, .
\end{eqnarray}
To derive Eq.~(\ref{bE1st}) we have used the fact that $\delta^{\rm gi(1)}({\bf q})$ can be neglected in Eq.~(\ref{a1}). 
We see that the use of a gauge-invariant mass density contrast valid at all scales introduces a scale dependence in the Eulerian bias $b_1^{\rm E}$ due to the relativistic effects which is not present 
if gauge-dependent density contrasts are used. Of course in the Newtonian limit $k\tau\gg 1$ one recovers the standard scale-independent Eulerian bias prediction from the Lagrangian approach, $b_1^{\rm E}\simeq (1+b_1^{\rm L})=(1+b^{\rm L}_{01}(M,\tau_{\rm f})/a(\tau))$.

At second order one first write the matter density contrast in Fourier space as

\begin{equation}
\label{defkerneldensity}
\frac{1}{2} \delta^{\gi(2)}(\k,\tau)=  \int \frac{d^3{\bf k}_1d^3{\bf k}_2}{(2 \pi)^3} \,{\cal K}^\gi_{\delta}({\bf k}_1,{\bf k}_2;\tau) \delta^{\gi(1)}({\bf k}_1,\tau) \delta^{\gi(1)}({\bf k}_2,\tau)
\delta_D ({\bf k}_{1}+{\bf k}_{2}-{\bf k}).
\end{equation}
With this position one finds 

\begin{eqnarray}
\label{defkerneldensityhalo}
\frac{1}{2} \delta^{\gi(2)}_{{\rm h}}(\k,\tau)&=&  \int \frac{d^3{\bf k}_1d^3{\bf k}_2}{(2 \pi)^3} \,b^{\rm E}_2({\bf k}_1,{\bf k}_2;\tau) \delta^{\gi(1)}({\bf k}_1,\tau) \delta^{\gi(1)}({\bf k}_2,\tau)
\delta_D ({\bf k}_{1}+{\bf k}_{2}-{\bf k}),\\
b^{\rm E}_{2}({\bf k}_1,{\bf k}_2;\tau)&=&\frac{b_1^{\rm L}(\k_1,\tau)}{2}+
\frac{b_1^{\rm L}(\k_2,\tau)}{2}+
\frac{1}{2}b_2^{\rm L}(\k_1,\k_2,\tau)\nonumber\\
&+&
\label{nl}
\frac{1}{2}b_1^{\rm L}(\k_1,\tau)\frac{f_\gi(\k_2,\tau_0)}{f_\gi(\k_2,\tau)a(\tau)}\frac{(\k_1\cdot\k_2)}{f_\gi(\k_1,\tau_0)k_1^2}+\frac{1}{2}b_1^{\rm L}(\k_2,\tau)\frac{f_\gi(\k_1,\tau_0)}{f_\gi(\k_1,\tau)a(\tau)}\frac{(\k_1\cdot\k_2)}{f_\gi(\k_2,\tau_0)k_2^2}\nonumber\\
&+&{\cal K}_{\delta}({\bf k}_1,{\bf k}_2;\tau),\nonumber\\
&&\\
b_2^{\rm L}(\k_1,\k_2,\tau)&\simeq&\frac{f_\gi(\k_1,\tau_0)}{f_\gi(\k_1,\tau)}\frac{f_\gi(\k_2,\tau_0)}{f_\gi(\k_2,\tau)}\frac{b^{\rm L}_{02}(M,\tau_{\rm f})}{a^2(\tau)}.
\end{eqnarray}
Again,  the Eulerian bias $b_2^{\rm E}$ gets a scale-dependence due to relativistic effects which arises once a gauge-invariant definition of the density contrasts are adopted.   Furthermore, in the second line of Eq. (\ref{nl}) there appears non-local terms proportional to the first-order bias parameter; they arise from the first-order density field evaluated at the Lagrangian point $\q$ when expressed in terms of the Eulerian position $\x$ \cite{biasn}.
The kernel ${\cal K}^\gi_\delta$ is  conveniently computed in the Poisson gauge (rather than in the synchronous gauge). In the presence of large local 
non-Gaussianities, we can introduce a gravitational potential which,   at some initial epoch and  deep in matter domination, reads  
\begin{equation}
\label{fnlphiin}
\phi_{\rm in}=\phi^{(1)}_{\rm in}+f^{\rm loc}_{\rm NL}(\phi^{(1)2}_{\rm in}-\langle \phi^{(1)2}_{\rm in} \rangle)\; ,
\end{equation}
with the dimensionless non-linearity parameter $f^{\rm loc}_{\rm NL}$ setting the level of quadratic local NG. In the case of 
large local non-Gaussianities, $f^{\rm loc}_{\rm NL}\gg 1$, one finds
\cite{uslast1,uslast2}

\begin{eqnarray}
\label{delta2P}
\frac{1}{2}\delta^{(2)}&=&\frac{\tau^4}{252}\left[5\left( \nabla^2 \varphi\right)^2 
+2 \varphi^{,ij}\varphi_{,ij}+7\varphi^{,i}\nabla^2\varphi_{,i} \right] 
-f^{\rm loc}_{\rm NL}\frac{\tau^2}{6}\nabla^2\varphi,\nonumber\\
\frac{1}{2}v^{(2)i}&=&\frac{\tau^3}{18}\left(-\varphi^{,ij}\varphi_{,j}+\frac{6}{7} 
\Psi^{,i}\right)+f^{\rm loc}_{\rm NL}\frac{\tau}{3}\partial^i\varphi^2 ,
\end{eqnarray}
where 
$\nabla^2 \Psi\equiv-\frac{1}{2}[(\nabla^2 \varphi)^2-\varphi_{,ik}
\varphi^{,ik}]$. Inserting these expressions into Eq.  (\ref{zz}),  we obtain

\begin{eqnarray}
\label{KER}
{\cal K}^\gi_\delta({\bf k}_1,{\bf k}_2;\tau)&=&
\frac{5}{7}+\frac{2}{7}\frac{
\left({\bf k}_1 \cdot {\bf k}_2\right)^2}{k_1^2 k_2^2}
- \frac{18 i}{7} \frac{1}{k^2 \tau} \left[\left({\bf k}_1 \cdot {\bf n}\right)+
\left({\bf k}_2 \cdot {\bf n}\right)\right] \left[1-\frac{\left({\bf k}_1 \cdot {\bf k}_2\right)^2}{k_1^2 k_2^2}
\right] \nonumber\\
&+&6i\left[\frac{\left({\bf k}_1 \cdot {\bf n}\right)}{k_1^2\tau}+\frac{\left({\bf k}_2 \cdot {\bf n}\right)}{k_2^2\tau}\right]\left[\frac{8}{21}+\frac{2}{7}\frac{
\left({\bf k}_1 \cdot {\bf k}_2\right)^2}{k_1^2 k_2^2}\right]
+6 f^{\rm loc}_{\rm NL}\frac{k^2}{k_1^2 k_2^2\tau^2} \nonumber \\
&-&18 i f^{\rm loc}_{\rm NL}\frac{\left[\left({\bf k}_1 \cdot {\bf n}\right)+
\left({\bf k}_2 \cdot {\bf n}\right)\right]}{k_1^2 k_2^2\tau^3} \, ,
\end{eqnarray}
where $k=\left|\k_1+\k_2\right|$ and we have performed an expansion in $(k_i\tau)^{-1}\ll 1 $ ($i=1,2$).
Notice that in the kernel the primordial non-Gaussian piece coming from the second-order density contrast is post-Newtonian and is damped 
by two powers of $(k_i\tau)$  with respect to the Newtonian leading terms. The Newtonian 
part of the kernel does not coincide with the one of the matter density contrast in the Poisson gauge
which can be found in Ref. \cite{uslast1}. Indeed, it gets a correction arising
 from the last term of the expression (\ref{zz}). 
Furthermore, 
there are  terms which are damped by only one power of $(k_i\tau)$; they originate  from  
the velocity contributions in gauge-invariant definition of the matter density contrast and they are absent 
if a gauge-dependent definition of the matter density contrast is adopted. The same holds for the last contribution, damped by three powers of  $(k_i\tau)$. 
This term comes from the primordial NG term in $v^{(2)}$ (which gives the dominant contribution to the second-order redshift perturbation~(\ref{z2})). 
All  other relativistic effects have been neglected.

 The Lagrangian bias factors $b_{01}^{\rm L}(\tau|M,\tau_{\rm f})$ and $b_{02}^{\rm L}(\tau|M,\tau_{\rm f})$ are those computed in Refs. \cite{biasp} (see also Eqs. (16) and (17) of 
 Ref. \cite{biasn}) through the extended  Press-Schechter approach and the peak-background split method (of course one can use 
 refinements of the Press-Schechter halo mass function like the one in Ref. \cite{ST}).
There are two key points in the Press-Schechter theory. On one side, the comoving number density of collapsed haloes is computed from the statistical 
properties of the linear density field, assumed to be Gaussian. On the other side, a patch of fluid is part of a collapsed region of radius $R$ if the smoothed linear 
density contrast on that scale exceeds a suitable threshold value $\delta_{\rm f}$ computed at the formation time $\tau_{\rm f}$.  In our gauge-invariant formulation 
the first point remains of course true; the second point requires though some comments. Indeed, the threshold value
for the matter density contrast is computed  according to the spherical collapse model. As we have commented, the latter requires to work in  the comoving-orthogonal gauge and therefore
$\delta_{\rm f}\equiv \delta_c\left(\tau_{\rm f}\right) \simeq 1.68/a(\tau_{\rm f})$ corresponds to the threshold matter density contrast in that gauge. This does not coincide with the 
gauge-invariant threshold density contrast.   However, it is easy to convince oneself that the gauge-invariant threshold value obtainable through the relation (\ref{d1})
differs from the one in the comoving-orthogonal gauge by a factor $10^{-4}$ even for very massive  haloes, $M\sim 10^{15} M_\odot$ (see also the discussion in Ref. \cite{SW}).

Adopting a gauge-invariant expression for the density contrast brings two new and interesting features in Eq.~(\ref{nl}) for  the Eulerian bias $b_2^{\rm E}$. 
First a new scale dependence parametrized by $f_{\rm gi}({\bf k}_i, \tau)$, accompanied by a characteristic kernel $K_\delta^{\rm gi}$. 
Second, the primordial NG introduces a dependence on the line of sight ${\bf n}$ which comes from terms like 
$\left({\bf v} \cdot {\bf n}\right)$, which are necessary to realize  the gauge-invariant 
definition of the matter density contrast. Therefore in the Eulerian bias there appears scale-dependent contributions which get also modulated. Notice that such 
${\bf v} \cdot {\bf n}$ contributions do not represent the usual redshift-space distortion effects, but rather they appear as effective corrections to the Eulerian bias 
Obviously, when computing the power spectrum of two objects with the same bias on large scales, this modulation disappears being the power specturm
real. Nevertheless, this does not happen
when computing the power spectrum of two objects with different bias.
In such a case, the resulting bias in the presence of some primordial NG is not only scale-dependent \cite{Dalal}, but also
depending on the angles $\cos\theta=\hat{{\bf k}}\cdot \hat{\bf n}$ between the vector ${\bf k}$ and the vector indicating the
 line-of-sight. A computation similar to the one leading the a scale-dependent bias when some large local NG is included  leads to a correction
to the bias  on large scales for two objects with different bias $b_{1\rm a}^{\rm E}(k)$ and  $b_{1\rm b}^{\rm E}(k)$

\begin{equation}
\Delta b_{1}^{\rm E}(k,R_1,R_2)=9if_{\rm NL}^{\rm loc}\delta_c(z)\frac{\Omega_{\rm m} H_0^2}{k^2T(k)}
\frac{H(z)f(\Omega_{\rm m})}{(1+z)k}
\left(b_{1\rm a}^{\rm E}(k,R_1)\frac{k^2\sigma_v^2(R_1)}{\sigma^2(R_1)}-b_{1\rm b}^{\rm E}(k,R_2)\frac{k^2\sigma_v^2(R_2)}{\sigma^2(R_2)}\right)\cos\theta\, ,
\end{equation}
where we have generalized our computation to a $\Lambda$CDM model,  $\Omega_{\rm m}$ is the dark matter critical density,
 $T(k)$ is the linear transfer function, $f(\Omega_{\rm m})\simeq \Omega_{\rm m}^{0.6}$ and $\sigma^2(R)$ and $\sigma^2_v(R)$ are the variance of the 
density contrast and of the velocity at a radius $R$, respectively.

\section{Conclusions}
\noindent 
In this paper we have described the computation of the Eulerian bias at second-order in perturbation theory. 
Paying attention to the gauge-invariant issues which necessarily arise when dealing 
with relativistic effects on large scales and with real observables, we have shown that some interesting effects 
show up. First 
of all, the Eulerian bias acquires a scale-dependence on large scales even if the primordial NG is totally
negligible. Secondly, the primordial non-Gaussianity induces in the bias  a modulation with the line of 
observation when sources
with different biases are observed. Of course, we are well aware of the fact that our results are not complete in the sense that 
not all effects have been included. In this paper we have restricted ourselves to that part of the
observed galaxy density contrast which is directly proportional to the dark matter density contrast through
the bias. On the contrary, 
we have not discussed, for instance, the redshift-space distortion and the magnification effects.
We leave it for future work.

\section*{Acknowledgments}
\noindent
A.R. acknowledges partial support by the EU Marie Curie Network UniverseNet (HPRNCT2006035863).
This research has been partially supported by the ASI/INAF
Agreement I/072/09/0 for the Planck LFI Activity of Phase E2. 

\bibliographystyle{JHEP}

\end{document}